\documentclass[apsrev,superscriptaddress,showpacs,showkeys]{revtex4}

\begin{document}
\title{Supersymmetry, PT-symmetry and Spectral Bifurcation}
\author{Kumar Abhinav}
\affiliation{Indian Institute of Science Education and Research-Kolkata, Mohanpur Campus, Nadia-741252, West Bengal, India}
\email{kumarabhinav@iiserkol.ac.in}
\author{P. K. Panigrahi}
\affiliation{Indian Institute of Science Education and Research-Kolkata, Mohanpur Campus, Nadia-741252, West Bengal, India}
\affiliation{Physical Research Laboratory, Navrangpura, Ahmedabad-380009, Gujarat, India}
\email{prasanta@prl.res.in}

\begin{abstract} 
We demonstrate that large class of PT-symmetric complex potentials, which can have isospectral real partner potentials, possess two different superpotentials. In the parameter domain, where the superpotential is unique, the spectrum is real and shape-invariant, leading to translational shift in a suitable parameter by \textit{real} units. The case of two different superpotentials, leading to same potential, yields broken PT-symmetry, the energy spectra in the two phases being separated by a bifurcation. Interestingly, these two superpotentials generate the two disjoint sectors of the Hilbert space. In the broken case, shape invariance produces \textit{complex} parametric shifts.
\end{abstract}



\maketitle

\section{Introduction}
The fact that two different superpotentials can give rise to same potential in supersymmetric quantum mechanics (SUSY-QM) is well-known \cite{Cooper}. The superpotential is deformed by an additive function, which leads to the construction of isospectral Hamiltonians having a number of physical applications. Complex potentials, particularly the ones having parity (P) and time reversal (T) invariance have attracted significant attention in the recent literature \cite{Bender}. Appearance of real eigenvalues in certain parameter domain for these complex PT-symmetric potentials, have led to considerable interest in this class of not so well-studied spectral problems. It has been found that the spectrum is real, when the wavefunction respects PT-symmetry, whereas complex eigenvalues, paired by complex conjugation are realized when the wavefunctions do not respect the above symmetry. As a function of certain potential parameter(s), the energy eigenvalues show bifurcation, when the spectra transits from real to complex values. Detailed discussions and the progress in this field can be found in the following references \cite{Mos} and \cite{Pramana}. A number of models have been studied, both numerically and analytically for illustrating the above structure in \cite{K}, and subsequently utilised in \cite{Ranjani}. It was observed that complex PT-symmetric potentials, under suitable parametrization, can be isospectral to real potentials \cite{Bagchi}. A generalized P\"oschl-Teller-type complex potential, of the above form, has been constructed by Ahmed \cite{Ahmed}, which reveals all the above mentioned features explicitly. Interestingly, it was found that SUSY only yields the real spectra, where wave-functions respect PT-symmetry. Explicit solutions of differential equation led to the complex branch of the spectra under different parametric conditions, where the solutions do not respect PT-symmetry. It is then natural to ask, what is the precise relationship between SUSY and PT-symmetry and if the aforementioned result is generic.
\\
\\
In this paper, it is shown that the presence of complex parameters in the superpotential can lead to the realization of the PT-symmetric complex potential through two different superpotentials. As compared to the case of isospectral deformation, in the present case the superpotentials do not differ by an additive function. The difference in superpotentials appear through the complex constant parameters.  In the broken PT phase, two superpotentials are present leading to the same potential, whereas in the unbroken phase the correspondence between the potential and superpotential is unique. This is possible only for complex potentials and does not manifest in the case of the real ones. The two different parameter domains explain the observed bifurcation separating the unbroken PT case from the broken one. In the latter phase, shape-invariance leads to complex translational shifts in the relevant parameters, whereas in the former case the translation is in real units \cite{Dabrowwaska}. Very interestingly, in case of complex eigenvalues, half of the Hilbert space is generated by one superpotential, whereas the other one generates the remaining part of the wavwfunctions.
\\
\\
The paper is organized as follows. In the following, we provide a brief introduction to SUSY-QM and proceed to construct the potentials from a general superpotential, in section III, starting from Ahmed's potential. After illustrating the nature of PT-symmetry in the broken sector, we derive the spectra, in both broken and unbroken phases, through shape-invariance. We then list examples of a large class of potentials exhibiting the above characteristics. In section IV, a number of solvable complex potentials are constructed through the above procedure, which are not PT-symmetric. These shape-invariant potentials constitute \textit{complexified} P\"oschl-Teller-type \cite{Morse} and Coulomb-type potentials, which undergo complex translation in the appropriate parameter domain. Finally, we conclude after pointing out a number of interesting directions in which the present investigation can be advanced.
\\
\\

\section{Supersymmetric Quantum Mechanics}
Supersymmetric quantum mechanics \cite{Cooper,Witten,Dutt,Darboux} interrelates the spectra of two different Hamiltonians, $H_{\pm}$, which can be written in the factorized form \cite{D}, where $ H_- (x)=A^{\dagger}A$ and $ H_+ (x)=AA^{\dagger}$ ($\hbar=1=2m$):
\begin{eqnarray*}
H_\pm (x)=-\frac{\partial^2}{\partial x^2} + V_\pm (x).
\end{eqnarray*}
\\
In terms of the superpotential W(x), the potentials are
\begin{equation}
V_\pm =W^2(x)\pm \frac{\partial W(x)}{\partial x};
\end{equation}
\\
here,
\begin{eqnarray*}
A^{\dagger}=\left(-\frac{\partial}{\partial x}+W(x)\right),\\     
A=\left(\frac{\partial}{\partial x}+W(x)\right),\\      \textrm{\,\,\, and \,\,\,}
W(x)=-\frac{1}{\psi_0(x)}\frac{\partial\psi_0 (x)}{\partial x}; 
\end{eqnarray*}
\\
$\psi_0 (x)$ being the ground-state eigenfunction of $H_{-}(x)$.
\\
\\
It can be straightforwardly shown that, if $\psi^{-}_{n} (x)$ is an eigenfunction of $H_-$, $A\psi^{-}_{n}(x)$ is an eigenfunction of $H_+$ with the same eigenvalue, except for the ground state of $H_-$ defined as $A\psi_0(x)=0$. Similarly if $\psi^{+}_{n} (x)$ is an eigenfunction of $H_+$, then $A^{\dagger}\psi^{-}_{n} (x)$ is an eigenfunction of $H_-$. Taking into account the unpaired ground state $\psi_{0}(x)$, the respective eigenvalues are related as,
\begin{eqnarray*}
E^{+}_{n}=E^{-}_{n+1}.
\end{eqnarray*}
\\
The corresponding eigenfunctions satisfy,
\begin{eqnarray*}
\psi^{+}_{n}(x)=[E^{-}_{n+1}]^{-\frac{1}{2}}A\psi^{-}_{n+1},\\     \textrm{\,\,\, and \,\,\,}     
\psi^{-}_{n+1}(x)=[E^{+}_{n}]^{-\frac{1}{2}}A^{\dagger}\psi^{-}_{n}.
\end{eqnarray*}
\\
 It was shown by Gendenshtein \cite{Gendenshtein} that if two isospectral potentials satisfy the relation
\begin{equation}
V_{+}(x;a_{0})=V_{-}(x;a_{1})+R(a_{1}),
\end{equation}
\\
where $a_{0}$ is a set of parameters for a given pair $V_{\pm}$, $a_{1}=f(a_{0})$, and $R(a_{1})$ is independent of x, then one can construct a hierarchy of Hamiltonians,
\begin{eqnarray*}
H^{s} &=& -\frac{\partial^2}{\partial x^2}+V_{-}(x;a_{s})+\sum_{k=1}^{s}R(a_{k})\\
      &=& -\frac{\partial^2}{\partial x^2}+V_{+}(x;a_{s-1})+\sum_{k=1}^{s-1}R(a_{k})
\end{eqnarray*}
\\
with ground  state energy
\begin{eqnarray*}
E_{0}^{s}=\sum_{k=1}^{s}R(a_{k}). 
\end{eqnarray*}
\\
On identifying that $H^{1}=H_{+}$ and $H^{0}=H_{-}$, it is found that the spectrum of $H_{-}$ is given as
\begin{equation}
E_{0}^{n}=\sum_{k=1}^{n}R(a_{k}).
\end{equation}
\\
Isospectral potentials satisfying Eq.(2) are called \textit{shape-invariant} \cite{Gendenshtein}. Using the relation $\psi_{n}^{-}(x;a_{0})\propto A^{\dagger}(x;a_{0})\psi_{n}^{-}(x;a_{1})$ it can be shown that
\begin{eqnarray*}
\psi_{n}^{-}(x;a_{0})\propto A^{\dagger}(x;a_{0})A^{\dagger}(x;a_{1})..............A^{\dagger}(x;a_{n-1})\psi_{0}^{-}(x;a_{n}).
\end{eqnarray*} 
\\
Therefore for shape-invariant potentials, one can solve the eigenvalue problem purely algebraically and obtain the spectrum and corresponding eigenfunctions. Detailed analysis of the properties of PT-symmetric quantum mechanical systems exhibiting supersymmetry has been carried out by Mostafazadeh \cite{Ali} and Plyushchay et al. \cite{M}, including the construction of appropriate norm \cite{Ali}. The question of norm in general pseudo-Hermitian Hamiltonian systems has been answered recently \cite{Das}.

\section{Construction of Complex PT-Symmetric Potentials}
Complex PT-symmetric potentials are known to have real eigenvalues, despite not being Hermitian in the usual sense. Further, this realness of spectra survives only over a specific range of parameter values, beyond which PT-symmetry is broken. In the broken PT range one observes complex conjugate spectra, as a function of certain potential parameters, with the corresponding states connected by PT-operation. If such a potential is shape-invariant, SUSY can be applied to algebrically obtain the spectra and corresponding eigenfunctions.  
\\
\\
Here, we start with a general superpotential, which leads to \textit{both} real and complex spectra. Under cetain parametrization, the superpotential is \textit{unique}, leading to real eigenvalues and corresponding wave-functions. For a different parametrization, two \textit{different} superpotentials yield the \textit{same} potential, leading to complex conjugate eigenvalues with corresponding eigenfunctions related by PT-operation. As a result, we arrive at the \textbf{supersymmetric parameter condition} for \textit{phase-transition} of the spectrum from real to complex conjugate values owing to spontaneous breaking of PT-symmetry. The experimental observation of phase-transition in a PT-symmetric system has recently been reported by Guo et al. \cite{Guo}.
\\
\\
To start with, we consider Ahmed's potential \cite{Ahmed}, which is given as,
\begin{equation}
V(x)=-V_{1}sech^2(\alpha x)-iV_{2}sech(\alpha x)tanh(\alpha x).
\end{equation} 
\\
Explicit solution of the Schr\"odinger equation shows that it has both real and complex-conjugate spectra, depending upon the relation among the real parameters $V_{1}$ and $V_{2}$, given as  $\vert V_{2} \vert \leq V_{1}+\frac{1}{4}$ and $\vert V_{2} \vert > V_{1}+\frac{1}{4}$, respectively \cite{Ahmed}. 
\\
\\
In order to see the role of supersymmetry and PT-symmetry, we start with the complex superpotentials 
\begin{equation}
W_{PT}^{\pm}=(A\pm iC^{PT})tanh(\alpha x)+(\pm C^{PT}+iB)sech(\alpha x),
\end{equation}
\\
where $A, B, C^{PT}$ are real constant parameters. Instead of repeating $C^{PT}$ in both the coefficients above, in general, one can start with different parameters. We have chosen them to be same to arrive at Ahmed's potential. The general case will be discussed afterwards. The $V_{-}(x)$ corresponding to Eq.(5) is, 
\begin{eqnarray}
V_{-}^{\pm}(x)=-\left[(A\pm iC^{PT})(A\pm iC^{PT}+\alpha)-(\pm C^{PT}+iB)^{2}\right]sech^{2}(\alpha x)\nonumber\\
-i(\pm iC^{PT}-B)\left[2(A\pm iC^{PT})+\alpha \right]sech(\alpha x)tanh(\alpha x),
\end{eqnarray}
\\
which in general may not be PT symmetric. To be PT-symmetric, the coefficient of the first term must be real and that of the second term must be purely imaginary. Implementing these conditions, one arrives at a unique relation:
\begin{equation}
C^{PT}\left[2(A-B)+\alpha \right]=0.
\end{equation}
\\

This is the most important result of the present paper, which we now study in more detail.
\\
\\

\textbf{\textit{Condition.1}}: If $C^{PT}=0$, then
\begin{equation}
W_{PT}(x)\equiv W_{real}(x) =Atanh(\alpha x)+iBsech(\alpha x),
\end{equation}
\\
which gives
\begin{equation}
V_{-}^{\pm}(x)\equiv V_{-}(x)=-\left(A(A+\alpha)+B^2\right)sech^2(\alpha x)+i\left(B(2A+\alpha)\right)sech(\alpha x)tanh(\alpha x).
\end{equation}
\\
This matches with Ahmed's potential, with the identifications,
\begin{eqnarray*}
V_{1}&=& A(A+\alpha)+B^2,\\   \textrm{\,\,\, and \,\,\,}
V_{2}&=&-B(2A+\alpha),
\end{eqnarray*}
\\
and satisfiy the reality condition $\vert V_{2} \vert \leq V_{1}+\frac{1}{4}$. From Eqs. (1),(2), (3) and (9) we get the corresponding known spectrum \cite{Dutt} 
\begin{equation}
E_{n}=-(n\alpha -A)^{2},
\end{equation}
\\
modulo a constant term. One can also obtain the corresponding eigenfunctions:
\begin{eqnarray}
\psi _{n}(x)\propto \left(sech(\alpha x)\right)^{\frac{A}{\alpha}} \exp [ -i\frac{B}{\alpha} tan^{-1}\left(sinh(\alpha x)\right)]P_{n}^{-\frac{A}{\alpha}-\frac{B}{\alpha}-\frac{1}{2}, -\frac{A}{\alpha}+\frac{B}{\alpha}-\frac{1}{2}} \left[i sinh(\alpha x)\right],
\end{eqnarray} 
\\
using Eq.(4) as per the prescription in \cite{Dabrowwaska}, which matches with those given in \cite{Ahmed} for real spectrum.
\\
\\

\textbf{\textit{Condition.2}}: For $C^{PT}\neq 0$, from Eq.(7), one gets $A=B-\frac{\alpha}{2}$, which when substituted in Eqs. (5) and (6) yields
 \begin{equation}
W_{PT}^{\pm}(x) \equiv W_{c}^{\pm}(x)=\left(A\pm iC^{PT}\right)tanh(\alpha x)+\left[\pm C^{PT}+i\left(A+\frac{\alpha}{2}\right)\right]sech(\alpha x)
\end{equation}
\\
and
\begin{eqnarray}
V_{-}^{\pm}(x) \equiv V_{-}^{c}(x)=-\left[2A(A+\alpha)-2(C^{PT})^{2}+\frac{{\alpha}^{2}}{4}\right]sech^{2}(\alpha x)\nonumber\\
+i\left[2A(A+\alpha)+2(C^{PT})^{2}+\frac{{\alpha}^{2}}{2}\right]sech(\alpha x)tanh(\alpha x).
\end{eqnarray}
\\
From the above we get
\begin{eqnarray*}
V_{1}&=& 2A(A+\alpha)-2(C^{PT})^{2}+\frac{{\alpha}^{2}}{4},\\    \textrm{\,\,\, and \,\,\,}
V_{2}&=&-\left[2A(A+\alpha)+2(C^{PT})^{2}+\frac{{\alpha}^{2}}{2}\right];
\end{eqnarray*}
\\
which satisfiy the condition $\vert V_{2} \vert > V_{1}+\frac{1}{4}$.  It needs to be emphasized that \textit{both} the superpotentials in Eq.(5) leads to the \textit{same} PT-symmetric complex potential, corresponding to broken PT symmetry. We can obtain their spectrum through shape-invariance,
\begin{eqnarray}
E_{n}^{\pm}=2n\left(A\pm iC^{PT}\right)\alpha + (n\alpha)^{2},
\end{eqnarray}
\\
which has complex conjugate pairs and each member of a complex conjugate pair can be connected to one of the two superpotentials in Eq.(12). Same can be said about the corresponding eigenfunctions,
\begin{eqnarray}
\psi _{n}^{\pm}(x)\propto \left(sech(\alpha x)\right)^{\frac{1}{\alpha}\left(A \pm iC^{PT}\right)}\exp \left[ \left(-\frac{i}{\alpha}\left(A+\frac{\alpha}{2}\right)\mp \frac{C^{PT}}{\alpha}\right)tan^{-1}\left(sinh(\alpha x)\right)\right]\nonumber\\
P_{n}^{\mp i2\frac{C^{PT}}{\alpha}, 2\frac{A}{\alpha}+\frac{1}{2}} \left[i sinh(\alpha x)\right],
\end{eqnarray} 
\\
which match with the previous results \cite{Ahmed}, under SUSY parametrization. They are related to each-other through PT transformation.
\\
\\
As mentioned earlier, one can start with two different parameters, say $C^{PT}$ and $D^{PT}$. Then we arrive at \textit{two} conditions,
\begin{eqnarray*}
D^{PT}&=& \frac{2BC^{PT}}{2A+\alpha},\\     \textrm{\,\,\, and \,\,\,}      
\left[(2A+\alpha)^{2}-(2B)^{2}\right]C^{PT} &=&0,
\end{eqnarray*}
\\
 for the potential to be PT-symmetric. Then if $C^{PT}\neq 0$, we have $\pm (2A+\alpha ) =2B$ and $D^{PT}=\pm C^{PT}$, which leads to the same results for the broken PT case. If $C^{PT}=0$, then $A$ and $B$ are unrestricted, giving the same results as in the unbroken PT case. 
\\
\\
The key observation about the potentials obeying the above property is that the odd parity part of the potential must have a purely imaginary coefficient and the even part must have a purely real one. With these constraints, one can construct a large class of PT-symmetric potentials, which are shape-invariant, starting from the known real shape-invariant potentials. Below, in table I, we list the superpotentials, the condition for which spectral bifurcation takes place leading from a PT-symmetric ground state to one which does not respect the same.
\\
\\
\\
\newpage
\textbf{Table I}: List of shape-invariant PT-symmetric potentials, with their respective superpotentials and parametric conditions for spectral bifurcation. The corresponding energies and ground state wave-functions are also shown. (Here we have taken $C^{PT}=C$ for simplicity.)
\\
\\
\begin{tabular}[c]{|c|c|c|c|c|c|}
\hline
$W(x;A,B,C)$      & Condition      & Cases       & $V_{-}(x)$     &Energy($E_{n}$)      &Ground-state \\
\hline 
$(A\pm iC)$   &$C(2A+\alpha)$     &$C=0$       &$-A(A+\alpha)sec^{2}(\alpha x)$     &$A^{2}-(A-n\alpha)^{2}$    &$\left(sech(\alpha x)\right)^{\frac{A}{\alpha}}exp(-i\frac{B}{A}x),$\\
$\times tanh(\alpha x)$   &$=0$    &     &$+2iBtanh(\alpha x)$   &$-\frac{B^{2}}{A^{2}}+\frac{B^{2}}{(A-n\alpha)^{2}}$    &$A>0$\\
\cline{3-6}
$+i\frac{B}{(A\pm iC)}$  &     &$C\neq 0,$     &$\left(A^{2}+C^{2}\right)sec^{2}(\alpha x)$    &$(A\pm iC)^{2}$   &$\left(sech(\alpha x)\right)^{\frac{1}{\alpha}(A\pm iC)}$\\
   &     &$A=-\frac{\alpha}{2}$   &$+2iBtanh(\alpha x)$    &$-(A\pm iC -n\alpha)^{2}$    &$\times exp\left(-\frac{B}{A^{2}+C^{2}}(iA\pm C)x\right)$\\ 
  &    &   &   &$-\frac{B^{2}}{(A\pm iC)^{2}}$    &$A>0$\\
    &    &   &   &$+\frac{B^{2}}{(A\pm iC -n\alpha)^{2}}$    &\\
\hline

$-(A\pm iC)$   &$C(2A-\alpha)$      &$C=0$      &$A(A-\alpha)csch^{2}(\alpha x)$   &$A^{2}-(A+n\alpha)^{2}$    &$\left(sinh(\alpha x)\right)^{\frac{A}{\alpha}}exp\left(-i\frac{B}{A}x\right),$\\
$\times coth(\alpha x)$   &$=0$    &    &$-2iBcoth(\alpha x)$    &$-\frac{B^{2}}{A^{2}}+\frac{B^{2}}{(A+n\alpha)^{2}}$     &$-\frac{1}{2\alpha}\leq A<0$\\
\cline{3-6}
$+i\frac{B}{(A\pm iC)}$   &    &$C\neq 0,$    &$-\left(A^{2}+C^{2}\right)csch^{2}(\alpha x)$      &$(A\pm iC)^{2}$
  &$\left(sinh(\alpha x)\right)^{\frac{1}{\alpha}(A\pm iC)}$\\            
    &    &$A=\frac{\alpha}{2}$     &$-2iBcoth(\alpha x)$      &$-(A\pm iC +n\alpha)^{2}$     &$\times exp\left(-\frac{B}{A^{2}+C^{2}}(iA\pm C)x\right),$\\ 
   &    &    &    &$-\frac{B^{2}}{(A\pm iC)^{2}}$    &$-\frac{1}{2\alpha}\leq A<0$\\
   &    &    &     &$+\frac{B^{2}}{(A\pm iC+n\alpha)^{2}}$         &\\
\hline

$(A\pm iC)$   &$C(2A-\alpha)$    &$C=0$       &$A(A-\alpha)sec^{2}(\alpha x)$    &$-A^{2}+(A+n\alpha)^{2}$   &$\left(cos(\alpha x)\right)^{\frac{A}{\alpha}}exp\left(-i\frac{B}{A}x\right)$\\
$\times tan(\alpha x)$    &$=0$   &    &$+2iBtan(\alpha x)$     &$-\frac{B^{2}}{A^{2}}+\frac{B^{2}}{(A+n\alpha)^{2}}$    &\\
\cline{3-6}
$+i\frac{B}{(A\pm iC)}$   &   &$C\neq 0,$     &$-(A^{2}+C^{2})sec^{2}(\alpha x)$   &$-(A\pm iC)^{2}$    &$\left(cos(\alpha x)\right)^{\frac{1}{\alpha}(A\pm iC)}$\\
     &      &$A=\frac{\alpha}{2}$   &$+2iBtan(\alpha x)$   &$+(A\pm iC+n\alpha)^{2}$   &$\times exp\left(-\frac{B}{A^{2}+C^{2}}(iA\pm C)x\right)$\\
   &   &   &   &$-\frac{B^{2}}{(A\pm iC)^{2}}$   &\\
   &    &   &   &$+\frac{B^{2}}{(A\pm iC+n\alpha)^{2}}$    &\\
\hline
$(A\pm iC)$   &$C(2A+\alpha)$     &$C=0$     &$A(A+\alpha)csc^{2}(\alpha x)$     &$-A^{2}+(A-n\alpha)^{2}$  &$\left(sin(\alpha x)\right)^{-\frac{A}{\alpha}}exp\left(-i\frac{B}{A}x\right),$\\
$\times cot(\alpha x)$    &$=0$    &   &$+2iBcot(\alpha x)$    &$-\frac{B^{2}}{A^{2}}+\frac{B^{2}}{(A-n\alpha)^{2}}$   &$0<A\leq \frac{1}{2\alpha}$\\
\cline{3-6}
$+i\frac{B}{(A\pm iC)}$     &     &$C\neq 0,$   &$-(A^{2}+C^{2})csc^{2}(\alpha x)$    &$-(A\pm iC)^{2}$      &$\left(sin(\alpha x)\right)^{-\frac{1}{\alpha}(A\pm iC)}$\\
   &    &$A=-\frac{\alpha}{2}$    &$+2iBcot(\alpha x)$      &$+(A\pm iC-n\alpha)^{2}$    &$\times exp\left(-\frac{B}{A^{2}+C^{2}}(iA\pm C)x\right),$\\
  &  &  &  &$-\frac{B^{2}}{(A\pm iC)^{2}}$      &$0<A\leq \frac{1}{2\alpha}$\\
  &   &   &   &$+\frac{B^{2}}{(A\pm iC-n\alpha)^{2}}$   &\\
\hline
\end{tabular}
\\
\\
\\
\\
The complex conjugate spectra appear due to spontaneous breaking of PT-symmetry, the specific parametrization condition $C^{PT}\neq 0$ can be identified as the SUSY criteria for broken PT. This is different from the analytic parameter criterion, where PT is unbroken over a range of parameters. In contrast, here we have a discreet condition $C^{PT}= 0$ for the same. Furthermore, broken PT results in the bifurcation of the corresponding Hilbert space in terms of two distinct superpotentials; the corresponding eigenfunctions map into each-other under PT operation. They correspond to a unique complex potential $V_{-}^{c}(x)$, which is still PT-symmetric as required. Apart from $\alpha$, the Ahmed potential has two independent parameters owing to the condition $A=B-\frac{\alpha}{2}$ , just like the corresponding potential for unbroken PT following $C^{PT}=0$. Same can be seen in case of all the potentials listed above. As mentioned earlier, the non-uniqueness of the superpotential is well known from isospectral deformation, which incorporates an additional function obeying Bernoulli's equation \cite{Dutt}. Here, the non-uniqueness arises \textit{parametrically}. 

\section{Non-PT-symmetric Potentials}
In the previous section, we have seen how a generic isospectral complex potential, under certain parametric constraints, can demonstrate a link between PT-symmetry and SUSY-QM. We now study shape-invariant complex potentials which need not have PT-symmetry. Since these potentials are shape-invariant, one can obtain their spectra analytically.
\\
\\
We start with a minimal \textit{complexification} of the well-known P\"oschl-Teller \cite{Morse} potential:
\begin{equation}
U(x)=U_{a}sech^{2}(\alpha x)+U_{b}csch^{2}(\alpha x),
\end{equation}
\\
by considering two superpotentials
\begin{eqnarray}
W_{1}(x)=Atanh(\alpha x)+iBcoth(\alpha x),\nonumber\\    \textrm{\,\,\, and \,\,\,}
W_{2}(x)=iAtanh(\alpha x)+Bcoth(\alpha x).
\end{eqnarray}
\\
The respective isospectral potentials are,
\begin{eqnarray}
U_{1}^{\pm}(x;A,B)=-A(A\mp \alpha)sech^{2}(\alpha x)-B(B\pm i\alpha)csch^{2}(\alpha x),\nonumber\\    \textrm{\,\,\, and \,\,\,}
U_{2}^{\pm}(x;A,B)=A(A\pm i\alpha)sech^{2}(\alpha x)+B(B\mp \alpha)csch^{2}(\alpha x).
\end{eqnarray}
\\
Both of them are shape-invariant since,
\begin{eqnarray}
U_{1}^{+}(x;A,B)=U_{1}^{-}(x;A-\alpha ,B+i\alpha)+(A+iB)^{2}-(A+iB-2)^{2},\nonumber\\    \textrm{\,\,\, and \,\,\,}
U_{2}^{+}(x;A,B)=U_{2}^{-}(x;A+i\alpha ,B-\alpha)+(iA+B)^{2}-(iA+B-2)^{2}.
\end{eqnarray}
\\
This immediately yields their respective spectra:
\begin{eqnarray}
E_{n}^{1}=4n\alpha(A-n\alpha)+i4Bn\alpha , \nonumber\\     \textrm{\,\,\, and \,\,\,}
E_{n}^{2}=4n\alpha(B-n\alpha)+i4An\alpha .
\end{eqnarray}
\\
The separations between two successive energy levels $\bigtriangleup E_{n}=E_{n}-E_{n-1}$ are given respectively as 
\begin{eqnarray}
\bigtriangleup E_{n}^{1}=-4\alpha(A-2n\alpha-\alpha)+i2B\alpha , \nonumber\\    \textrm{\,\,\, and \,\,\,}
\bigtriangleup E_{n}^{2}=-4\alpha(B-2n\alpha-\alpha)+i2A\alpha .
\end{eqnarray}
\\
The corresponding eigenfunctions are,
\begin{eqnarray}
\psi_{n}^{1}(x)\propto \left[cosh(\alpha x)\right]^{-\frac{A}{\alpha}}\left[sinh(\alpha x)\right]^{-i\frac{B}{\alpha}}P_{n}^{(-i\frac{B}{\alpha}-\frac{1}{2},-\frac{A}{\alpha}-\frac{1}{2})}\left(sinh(\alpha x)\right),\nonumber\\   \textrm{\,\,\, and \,\,\,}
\psi_{n}^{2}(x)\propto \left[cosh(\alpha x)\right]^{-i\frac{A}{\alpha}}\left[sinh(\alpha x)\right]^{-\frac{B}{\alpha}}P_{n}^{(-\frac{B}{\alpha}-\frac{1}{2},-i\frac{A}{\alpha}-\frac{1}{2})}\left(sinh(\alpha x)\right).
\end{eqnarray}
\\
\\
It is straightforward to check that, when the above potential is real, one finds the corresponding energies to be $E_{n}^{real}=4n\alpha(A+B-n\alpha)$, with energy spacings $\bigtriangleup E_{n}^{real}=-4\alpha(A+B-2n\alpha-\alpha)$. The corresponding eigenfunctions are $\psi_{n}^{real}(x)\propto \left[cosh(\alpha x)\right]^{-\frac{A}{\alpha}}\left[sinh(\alpha x)\right]^{-\frac{B}{\alpha}}P_{n}^{(-\frac{B}{\alpha}-\frac{1}{2},-\frac{A}{\alpha}-\frac{1}{2})}\left(sinh(\alpha x)\right)$.
\\
\\
Due to the \textit{minimal complexification} the original real spectra turns into a complex spectra with the real part mimicking the original spectra. Interestingly, the imaginary part of the energy displays equispaced harmonic oscillator-like spectra. The non-uniqueness of the original spectra arising from infinite possible combinations of parameters A and B, giving the same (A+B), is also removed. As in both the examples of complexification, only the parameter that got multiplied with $i$ appears in the equispaced imaginary spectra, the complexification process does not differentiate between the two functions appearing in the superpotential.
\\
\\
From Eqs.(22), we see that the powers appearing in the wave-functions become imaginary. In case of the first wave-function in Eqs.(22), this completely removes the singularities of the function $sinh(\alpha x)$ for the whole range of B. Similarly, the singularity of the function $cosh(\alpha x)$ is removed in the second case for the whole range of A. Thus wave-functions for these \textit{complexified} potentials are normalizable over a greater range of parameters than those belonging to their real counterpart \cite{PKP}.
\\
\\

One can construct a complex Coulomb type potential starting with the superpotential,
\begin{equation}
W(r)=\frac{i\alpha}{r} +\beta,
\end{equation}
\\
which results in the potentials 
\begin{eqnarray*}
V_{\pm}(r;\alpha,\beta)=-\frac{\alpha (\alpha \pm i)}{r^{2}} + \frac{i2 \alpha \beta}{r} + \beta^{2}.
\end{eqnarray*}
\\
If the coefficient of $\frac{1}{r}$, $\alpha\beta$ is constant(say $e$), then one can compare $V_{-}$ with the radial Coulomb potential 
\begin{eqnarray*}
V_{C}(r)=\frac{l(l+1)}{r^{2}}-\frac{e}{r},
\end{eqnarray*} 
\\
where $l$ and $e$ carry the conventional meanings. The shape-invariance condition,
\begin{eqnarray*}
V_{+}(r;\alpha)=V_{-}(r;\alpha +i)+\gamma^{2}\left[{\frac{1}{{\alpha}^{2}}} - \frac{1}{(\alpha +i)^{2}}\right],
\end{eqnarray*}
\\
is satisfied and one finds,
\begin{equation}
E_{n}={\gamma}^{2} \left[\frac{1}{\alpha^{2}} - \frac{1}{(\alpha +in)^{2}}\right],
\end{equation}
\\
which is similer to the Hydrogen spectra. Starting from the superpotential in Eq.(23), the complex eigenfunctions can be obtained \cite{Dutt}. For example the ground-state wavefunction is given by
\begin{equation}
\psi_{0}(r)\propto r^{i\alpha}\exp^{-\beta r},
\end{equation}
which is well-behaved over the whole range of the real parameter $\alpha$, unlike the well known \textit{real} counterpart.
\\
\\

\section{Conclusions}
In conclusion, for a large class of potentials, it is found that a given PT-symmetric complex potential can be realized from two parametrically different superpotentials. In the parameter domain, where the superpotential is unique, the Hamiltonians yield real eigenvalues and the latter leads to the broken PT phase. In this case, the two superpotentials yield two disjoint parts of the Hilbert space of the Hamiltonian and shape-invariance leads to complex shifts in the appropriate potential parameters. Higher order supersymmetry has been applied to construct new classes of PT-symmetric potentials \cite{Q}, which may also reveal the features found here. Our procedure also yields non-PT-symmetric potentials with complex eigenvalues, which can be obtained through shape-invariance. This feature may manifest in many-body systems of the Calogero-Sutherland type \cite{Calogero,Gurappa,Kundu}. This dynamical system is currently under investigation. Finally, we have studied here translation-type shape-invariance. It is worth investigating, through our method complex supersymmetric potentials where shape invariance arises through scaling of the parameters. 
\\
\\
\textbf{Acknowledgment:} We acknowledge useful discussions with Prof. R. Dutt, and are thankful to Profs. A. Mostafazadeh and Plyushchay for bringing to our notice a number of related references.
\\
\\

\end{document}